\documentclass[twocolumn]{aastex631}

\usepackage{graphicx}	
\usepackage{amsmath}	

\usepackage{amssymb}	
\usepackage{longtable}
\usepackage{float}
\usepackage{gensymb}
\usepackage{color} 

\begin{document}

\title[Parsec-scale Properties of J1543-0757]{Two radio cores in GPS J1543-0757: a new dual supermassive black hole system?}

\correspondingauthor{Xiaopeng Cheng}
\email{xcheng0808@gmail.com}

\author[0000-0003-4407-9868]{Xiaopeng Cheng}
\affiliation{Korea Astronomy and Space Science Institute, 776 Daedeok-daero, Yuseong-gu, Daejeon 34055, Korea}

\author[0000-0002-4148-8378]{Bong Won Sohn}
\affiliation{Korea Astronomy and Space Science Institute, 776 Daedeok-daero, Yuseong-gu, Daejeon 34055, Korea}
\affiliation{Department of Astronomy and Space Science, University of Science and Technology, 217 Gajeong-ro, Daejeon, Korea}



\begin{abstract}

We report on the discovery of a dual supermassive black hole system in the radio galaxy J1543$-$0757, with a
projected separation between the two black holes of $\sim$ 46 mas. 
The result is based on recent multifrequency observations using the Very Long Baseline Array and European VLBI Network, which reveal two compact, variable, flat-spectrum, active nuclei within the elliptical host galaxy of J1543$-$0757. Multiepoch observations from the VLBA also provide constraints on the spectral index and proper motions of all components. 
The flat spectra of both N and S at both frequencies strongly support that these radio components are associated with two separate, accreting supermassive black holes (SMBHs). 
The two nuclei appear stationary, while the jets emanating from the weaker of the two nuclei appear to move out and terminate in bright hot spots. 
The discovery of this system has implications for the number of close dual black holes that might be sources of gravitational radiation.
\end{abstract}



\section{Introduction} \label{sec:intro}

There is strong observational evidence that almost massive galaxies in the universe have supermassive black holes (SMBHs) at their centers \citep{1995ARA&A..33..581K,2013ARA&A..51..511K}.
One of the possible scenarios for SMBHs formation is through a hierarchical merging of host galaxies based on the standard $\Lambda$CDM cosmology \citep{1978MNRAS.183..341W}. The merging galaxy is expected to result in dual SMBHs in the center of the merger product and the two SMBHs are typically to merge \citep{2003ApJ...582..559V}.
While the fact that galaxy mergers are not uncommon, direct observational evidence for dual SMBHs on parsec(pc)-scale separation remains surprisingly handful \citep[e.g.,][]{2006ApJ...646...49R,2011MNRAS.410.2113B,2014Natur.511...57D,2017NatAs...1..727K,2017MNRAS.471.1873Y,2022A&A...663A.139A}.
Searching dual SMBHs can help to enrich our understanding of a complete picture of galaxy merger evolution \citep{1980Natur.287..307B}.

Due to the small angular separation, very long baseline interferometry (VLBI) is the only technique that can identify double compact flat- or inverted-spectrum components from accreting dual SMBHs at projected separations of a milli-arcsecond (mas), which correspond to pc-scales for nearby active galactic nuclei (AGNs) \citep{1980Natur.287..307B,2003AIPC..686..161K,2018RaSc...53.1211A,2021ApJ...914...37B,2023ApJ...958...29C}.
However, compact symmetric objects (CSOs) can also show similar morphology ($<$ 1 kpc, double-lobed radio structure, slow relative proper motion) and inverted spectrum \citep{1994ApJ...425..568C,2012ApJS..198....5A,2012ApJ...760...77A,2014ApJ...792L...8W}.
The high-sensitivity and high-resolution VLBI observations are required to avoid false positives \citep[e.g.,][]{2012ApJ...745...67F,2014ApJ...792L...8W}.
Some recent VLBI observations have already detected double radio cores in several sources \citep[e.g.,][]{2006ApJ...646...49R,2017NatAs...1..727K,2017MNRAS.464L..70Y,2017MNRAS.471.1873Y}.
The tightest dual SMBHs candidate with a projected separation of 0.35 pc known by direct identification using the VLBI technique is NGC 7674 \citep{2017NatAs...1..727K}.
On pc-scales, the only imaged and well-studied dual SMBHs candidate is B3 0402+379 \citep{2006ApJ...646...49R,2017ApJ...843...14B}, with a projected separation of 7.3 pc.
Other identified dual SMBHs have projected separation over 100 pc \citep{2018Natur.563..214K,2020A&A...633A..79K,2023ApJ...942L..24K}.

\begin{figure}
    \centering
    \includegraphics[width=0.45\textwidth]{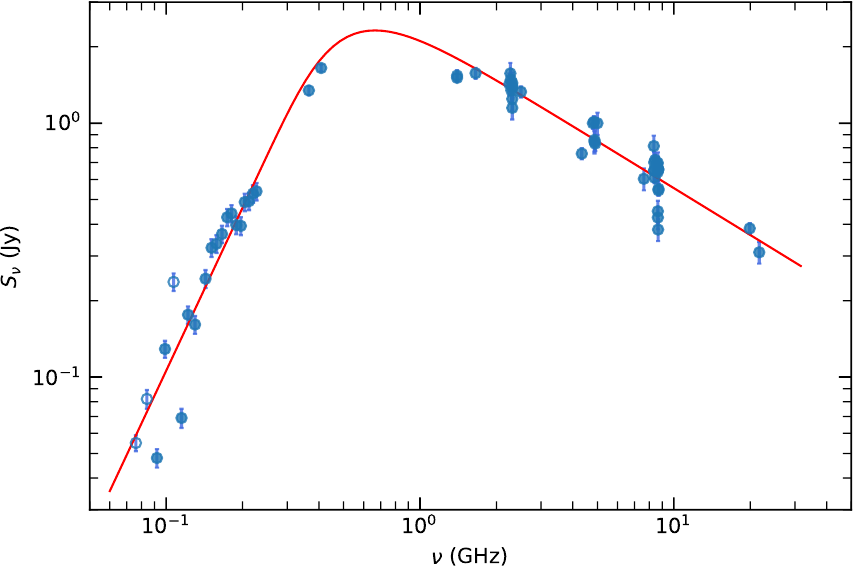}
    \caption{The radio spectra of J1543$-$0757. We used the data observed with single-dish radio telescopes and connected element interferometers, and the total flux density of VLBI observations in this paper which reflect the flux density of the entire source. The red curve represents the generic curved model of equation 3 in \citet{2017ApJ...836..174C}, with thick and thin spectral indices of $-$0.61$\pm$0.05 and 2.13$\pm$0.13, respectively.}
    \label{fig:sp}
\end{figure}

The radio galaxy J1543$-$0757 \citep[PKS 1540$-$077, z = 0.172;][]{2000A&AS..143..181D} is identified as a GPS source \citep{2000A&AS..143..181D,2004A&A...424...91E}.
The radio spectrum is constructed with data available from the NED\footnote{http://ned.ipac.caltech.edu/} (Table \ref{tab:SED}) and the total flux density of VLBI observations used in this paper (Table \ref{tab:observation log}), indicating a steep radio spectrum (Figure \ref{fig:sp}).
The frequencies shown in the x-axis have been corrected to the source rest frame.
A generic curved spectral model is adopted to fit the radio spectrum data \citep[e.g.,][]{2017ApJ...836..174C}, shown as a red-colored line.
The spectral index is $\alpha_{thin}$ = $-$0.61$\pm$0.05 at the optically thin part and $\alpha_{thick}$ = $-$2.13$\pm$0.13 at the optically thick part with a turnover frequency at about 0.51$\pm$0.04 GHz and peak flux density of 2.18$\pm$0.19 Jy, The peak flux density and turnover frequency are slightly higher than previous studies \citep{2002MNRAS.337..981S}.
We caution that these data are from non-simultaneous observations and different spatial frequency coverage.
The impact of systematic biases resulting from different spatial frequency coverage and no-simultaneous observations will be explored with future simultaneous multi-frequency observations. 
In fact, the total flux densities measured by non-simultaneous observations of almost thirty years are still roughly consistent with each other and can be well-fitted.

The source is unresolved in the 1.4 GHz NRAO VLA Sky Survey (NVSS) observations \citep{1998AJ....115.1693C}.
At pc-scales, the VLBI images of J1543$-$0757 display two extended components separated by $\sim$46 mas, with the northern component $\sim$3 three times brighter \citep{2000ApJS..131...95F}, similar to CSO-like radio morphology.
However, the pc-scale structure of J1543$-$0757 is unique in that it possesses two compact, flat-spectrum components and two pairs of radio jets, indicating two potential active nuclei.
To explore the nature of the source in more detail, we present the multi-frequency and multi-epoch VLBI observations of the radio galaxy J1543$-$0757.
There are three possibilities: 1) J1543$-$0757 is a CSO; 2) the nucleus was being gravitationally lensed; 3) two active nuclei of a dual supermassive black hole system in J1543$-$0757 with a projected separation of 46 mas.
To test these scenarios we investigated the available VLBI observations, designed to resolve any jet component and to look for relative motions and spectral indices.

Throughout the paper, we assume the standard $\Lambda$CDM cosmological model with $\rm H_{0}$ = 71 km\,s$^{-1}$\,Mpc$^{-1}$, $\Omega_{\rm M} = 0.27$, and $\Omega_{\Lambda} = 0.73$.
At a redshift of $z = 0.172$, an angular size of 46 mas corresponds to a projected linear size of 130 pc, and the conversion from angular velocity to projected linear speed is 1 mas yr$^{-1}$ = $9.445\, c$.

\section{Observations}
\label{sect:Obs}

\subsection{VLBA observations in 2019}
\label{VLBA in 2019}

The source was selected as the phase calibrator for Fermi J1544-0649 in our VLBA observations \citep[(project code: BT146, PI: P.H.T. Tam,][]{2022ApJ...934...39S}.
The observations were done on February 10 and May 20 2019 at \textbf{5} GHz, and on February 11 and May 21 2019 at \textbf{8} GHz.
The duration of each segment was about 1.5 hours.
We used a phase referencing technique with a 3.5-min cycle of ``J1543$-$0757 (1min)-J1544$-$0649 (2.5min)" at 5 GHz and a 1.5-min cycle of ``J1543$-$0757 (0.5min)-J1544$-$0649 (1min)" at 8 GHz.
The integration time is 21 minutes at 5 GHz and 22 minutes at 8 GHz. 
The data were recorded at a 2048 Mbps rate, with 2 polarizations, 2 intermediate frequency channels (IFs) per polarization, and 128 MHz bandwidth per IF.
Table \ref{tab:observation log} summarizes the  basic information of the observations.
The data were correlated using the DiFX software correlator (\citealt{2007PASP..119..318D,2011PASP..123..275D}) at Socorro with an averaging time of 2s, 256 frequency channels per IF and uniform weighting.

We calibrated the data in the US National Radio Astronomy Observatory (NRAO) Astronomical Imaging Processing System (AIPS, version 31DEC21; \citealt{2003ASSL..285..109G}).
\textit{A priori} amplitude calibration of the visibility was carried out using the system temperatures and antenna gains measured at each station during the observations.
The dispersive delays caused by the ionosphere were corrected according to a map of total electron content provided by Global Positioning System (GPS) satellite observations.
Phase errors due to the antenna parallactic angle variations were removed.
The instrumental single-band delays and phase offsets were corrected using 2-min observational data of the calibrator 3C 345.
After inspecting the data and flagging, global fringe-fitting was performed on the source J1543$-$0757 with a 0.5-min solution interval and a point-source model \citep{1995ASPC...82..189C} by averaging over all the IFs.
After fringe-fitting, the solutions for the source were applied to its own data by linear phase connection using rates to resolve phase ambiguities. 
In the final step, we used the task BPASS to calibrate the bandpass amplitude shapes by fitting a short data scan on a calibrator 3C345 and applied the solutions to all data.
Then we made single-source calibrated data sets with SPLAT/SPLIT and used the task FITTP to export the calibrated visibility data files to other workspaces.

A number of circular Gaussian components were fitted to the visibility data using the MODELFIT program in the Difmap software package (version 2.5e; \citealt{1997ASPC..125...77S}) to quantitatively describe the emission structure.
The model fitting parameters are presented in Table \ref{tab:model fitting}. The typical flux density uncertainty is about 10\%. 
The uncertainties of the fitted component size is less than 1/5 times the beam size. 
The separation of individual Gaussian components from
the core is consistent within 15\% of the fitted size of the Gaussian models.

\subsection{archival observations}

To further study this source, we searched and obtained the calibrated VLBI data from the European VLBI Network (EVN) data Archive\footnote{project code EB030, PI: H.E. Bignall, http://archive.jive.nl/scripts/portal.php}. The source was observed in a single epoch as a phase calibrator at 1.7 GHz with the EVN.
The integration time is about 24 minutes.
The correlated data was analyzed with AIPS and Difmap, following the standard VLBI data reduction procedures (the same procedure described in Section \ref{VLBA in 2019}).

We also searched the astrogeo database\footnote{VLBA calibrator survey database is maintained by Leonid Petrov, http://astrogeo.org/.}.
The source was observed 10 epochs simultaneously at dual frequencies (2 and 8 GHz, or 5 and 8 GHz) in snapshot mode from 1997 to 2018 with the VLBA, allowing us to investigate their structure change, variability, and spectral index which can be used to identify the flat-spectrum radio core.
The AIPS-calibrated data were directly obtained from the Astrogeo database. We performed only a few iterations of self-calibration in
Difmap software package to eliminate the residual amplitude and phase errors. After self-calibration, the visibility data were fitted with several circular Gaussian components to quantitatively describe the emission structure.
To reduce errors due to differences in data quality (in practice, the difference in data quality is not much), we used the same model fitting uncertainties for all VLBI data.
Further information regarding the observation can be found in Table \ref{tab:observation log}.

\section{Results}
\label{3}

\subsection{Radio morphology on parsec-scale}

\begin{figure*}
\centering
 \includegraphics[height=6cm]{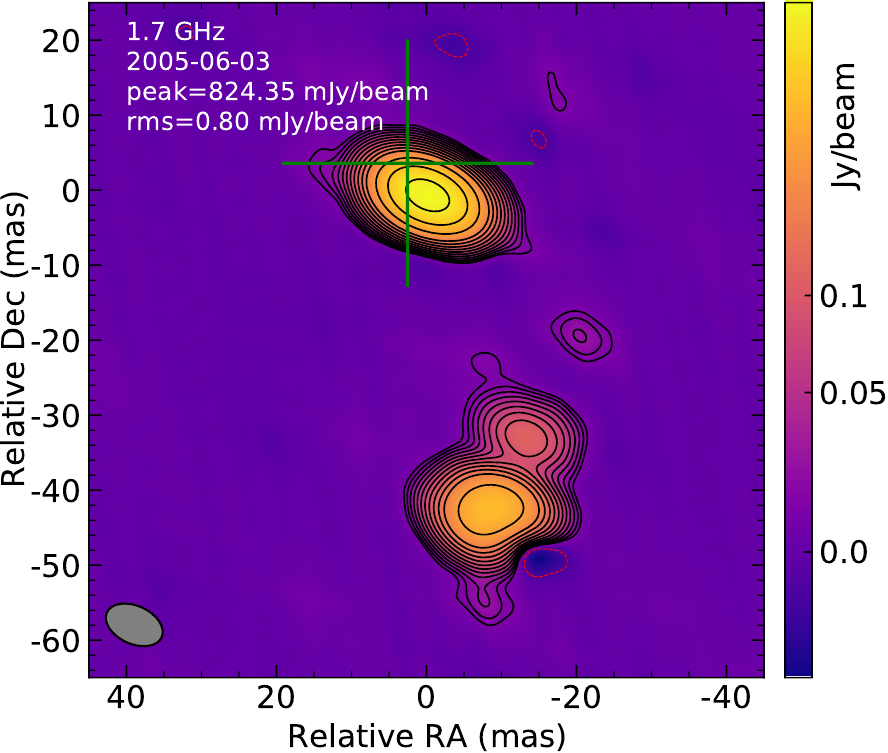}
 \includegraphics[height=6cm]{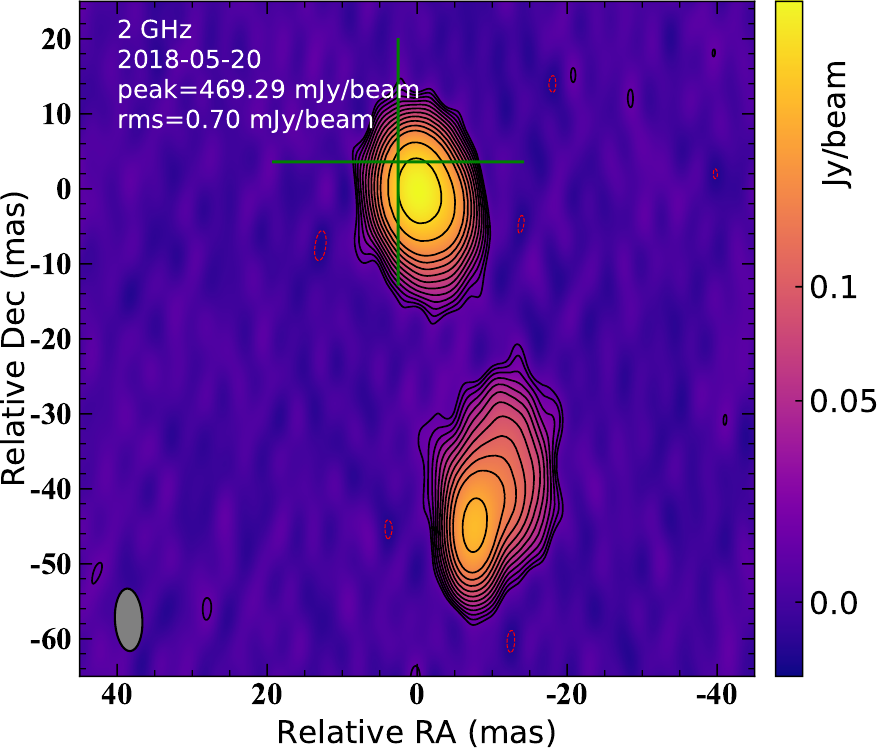}
\\
 \includegraphics[height=6cm]{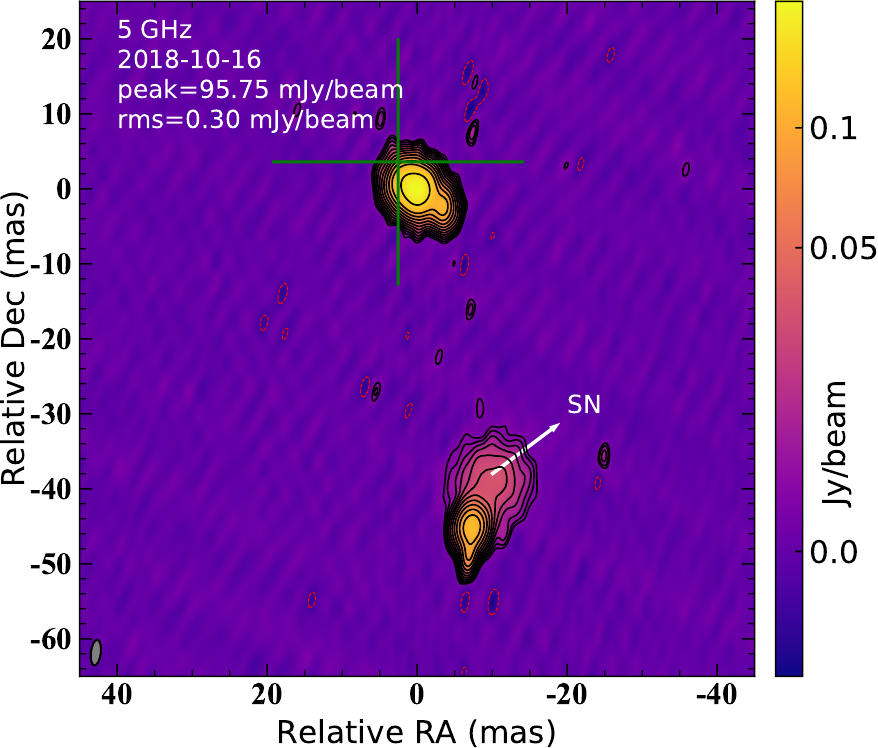}
 \includegraphics[height=6cm]{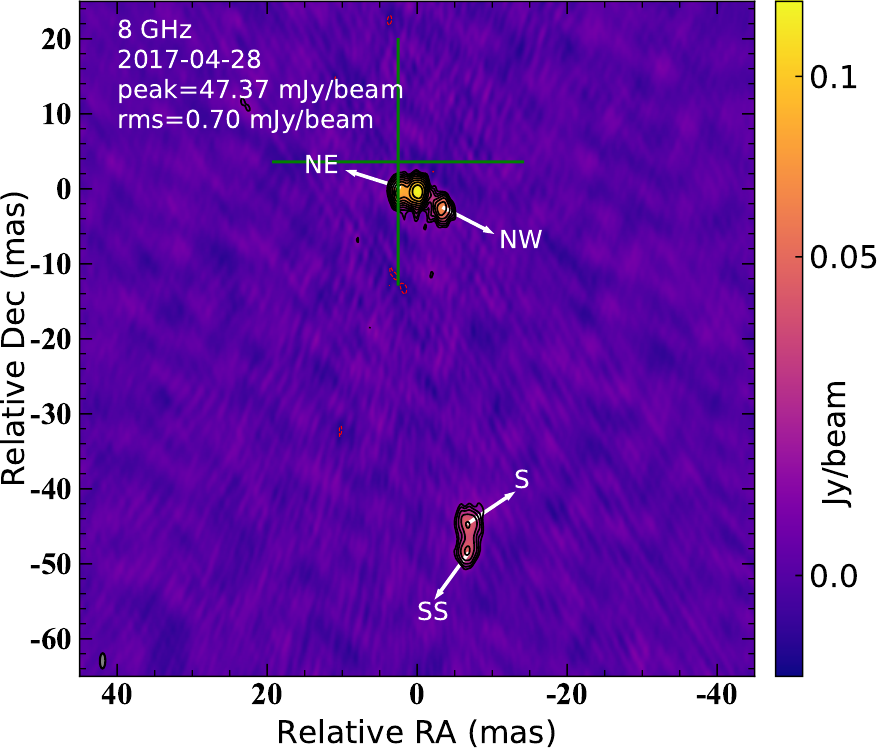}
\\
\caption{Naturally weighted total intensity VLBI images of J1543$-$0757 at 1.7 GHz, 2 GHz, 5 GHz, and 8 GHz. The peak intensity and rms noise are listed in upper-left corner of each panel. The lowest contours are at $\pm$3 times rms noise and further positive contours are drawn at increasing steps of 2. The grey-colored ellipse in the bottom-left corner of each panel denotes the restoring beam. The green-colored plus sign marks the position derived from the Gaia astrometry catalog \citep{2020yCat.1350....0G}, corresponding to the centroid of the optical emission of the AGN. The arrows represent the direction of motion found for each component, relative to the position of component N. Component N is at the center of the image.}
\label{fig:image}
\end{figure*}

Figure \ref{fig:image} shows the representative naturally weighted total intensity images of J1543$-$0757 observed with the EVN at 1.7 GHz and the VLBA at 2, 5, and 8 GHz.
The image size displayed is 90 $\times$ 90 mas$^{2}$ and the basic information is also listed in the images.
The optical Gaia position \citep[Gaia Early Data Release 3,][]{2020yCat.1350....0G} is also shown in Fig. \ref{fig:image}, marked as green pluses.
The source shows two discrete features: the north structure shows three components in the northeast to southwest direction, labeled as NE, N, and NW; the south structure shows a similar morphology with three main components in the southeast to northwest direction, labeled as SS, S, and SN.
The south and north structures are detected at a separation of $\sim$ 46 mas($\sim$ 130 pc) along PA = $-$170$\degree$.
The components NE and SS were not detected in the 1.7-GHz EVN map and 2-GHz VLBA maps, because they have about three times poorer resolution than the 5-GHz VLBA image.
In the high-resolution images observed at 5 and 8 GHz, the four main jet components NE, NW, SN, and SS are detected.
The arrows shown in Fig. \ref{fig:image} represent the direction of motion found for each component, relative to the position of component N (see Section \ref{3.3}).

To further study the complex structure, we fitted circular Gaussian models to the visibility data in Difmap and listed the best-fitted parameters in Table \ref{tab:model fitting}.
The component N is located at the center of the map and also near the Gaia position ($\sim$ 4.3 mas) with a mean size of $\sim$ 3.26$\pm$0.49 mas at 8 GHz and a mean value of the spectral index about $-$0.58 (column 8 of Table \ref{tab:model fitting}), indicating a flat radio spectrum of the core.
The component S, located at about 46 $\pm$ 2.5 mas at PA = $-$170$\degree$ from the core component N, shows a compact structure at both 5 and 8 GHz and has a flat radio spectrum with a mean value of the spectral index of $\alpha$ = $-$0.25 (column 8 of Table \ref{tab:model fitting}).
The compact size and flat radio spectrum suggest that component S is another self-absorbed core for J1543$-$0757.
The components NE, NW, and SN are detected in several epochs at 5 or 8 GHz.
Component SS is detected from 2017 April, suggesting a new jet component originating from the core and moving outwards.

\subsection{Variability}
\label{3.2}

\begin{figure}
    \centering
    \includegraphics[width=0.45\textwidth]{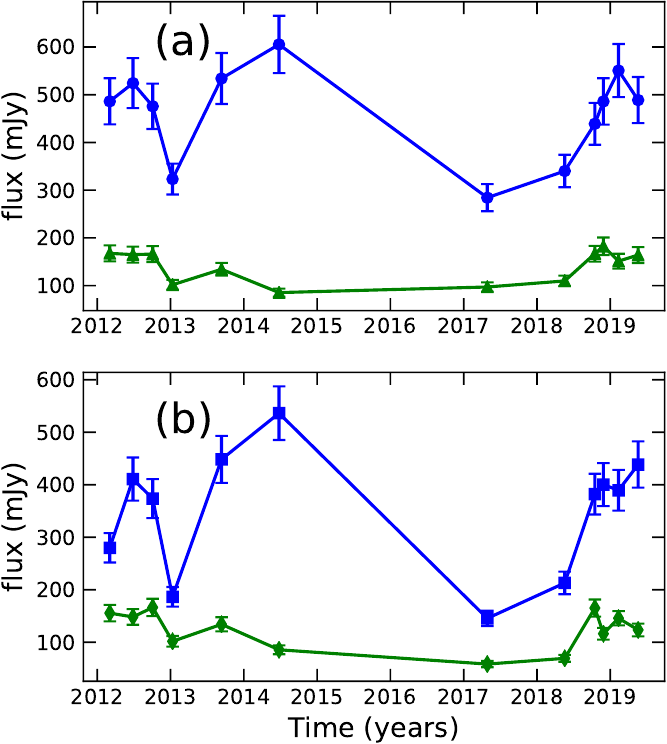}
    \caption{Light curves of the components of J1543$-$0757 at 8 GHz. (a) Total flux density of the north components N, NE and NW (blue, circle) and south components S, SN, and SS (green, triangle), (b) The flux density of the compact component N (blue, square) and S (green, diamond).}
    \label{fig:lightcurve}
\end{figure}

Figure \ref{fig:lightcurve}a shows the radio light curve at 8 GHz over a time span of 7 years, from 2012 February until 2019 May.
The upper panel shows the total flux density of northern (N, NE, and NW) and southern (S, SN, and SW) components, and Figure \ref{fig:lightcurve}b shows the light curve of components N and S.
The variability index, defined as $\frac{S_{max} - S_{min}}{S_{max} + S_{min}}$, is about 0.36 for northern components, and 0.42 for southern components. 
The total flux density of the northern components is approximately three times higher than the southern components, which is in good agreement with the previous result \citep{2004A&A...424...91E} and the south components have slightly larger variability.

The variability index is about 0.52 for component N and 0.48 for component S. 
The north component N shows more variability than component S. The variability is also different between S and N, which are shown in Fig. \ref{fig:lightcurve} and Table \ref{tab:model fitting}.
We note that the component NW exhibits very large variability exceeding 60\%, contributing slightly higher variability to the total flux density in northern components compared to southern components.

\subsection{Component proper motions }
\label{3.3}
\begin{figure*}
\centering
 \includegraphics[height=4.5cm]{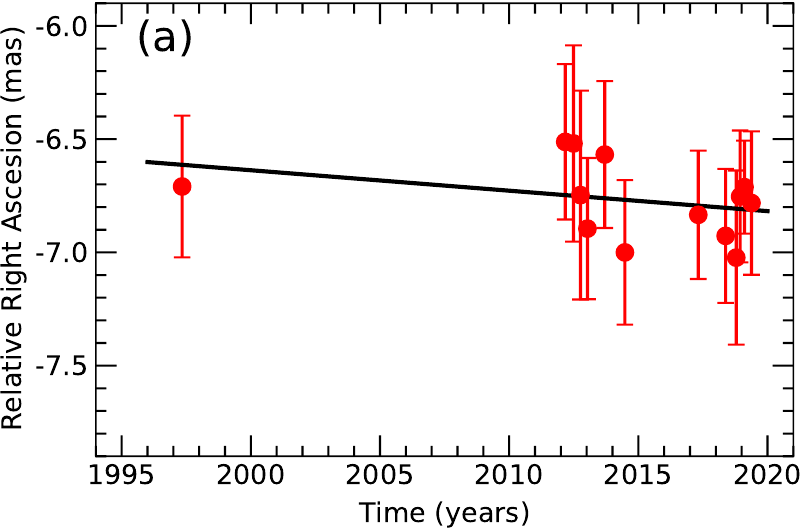}
 \includegraphics[height=4.5cm]{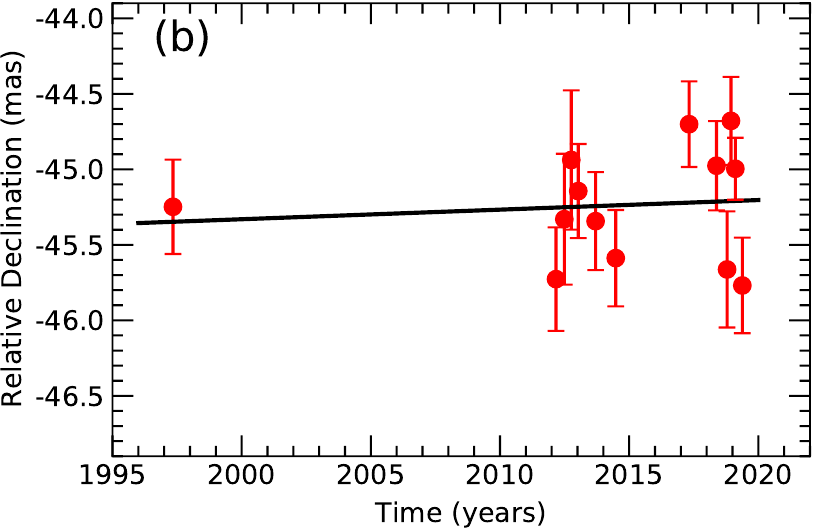}
\\
 \includegraphics[height=4.5cm]{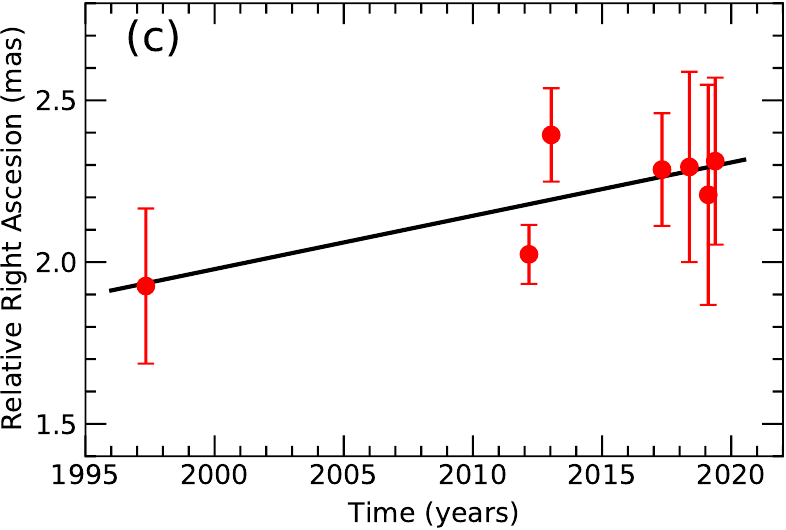}
 \includegraphics[height=4.5cm]{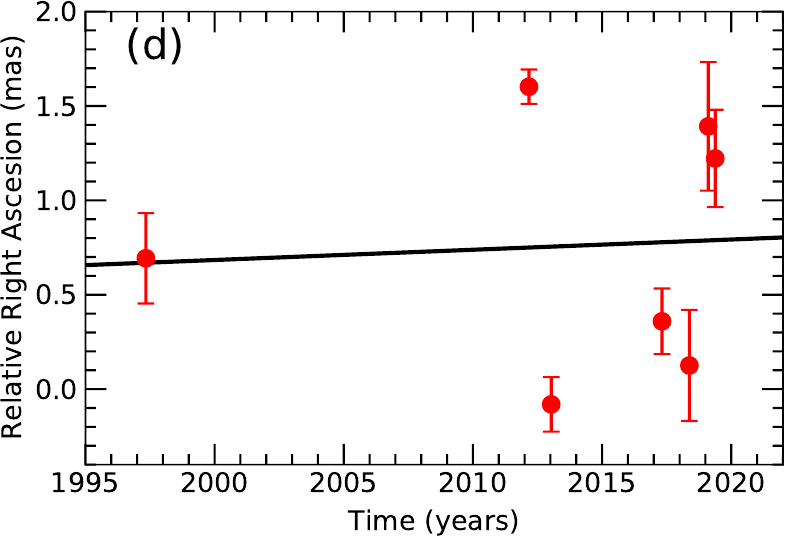}
\\
 \includegraphics[height=4.5cm]{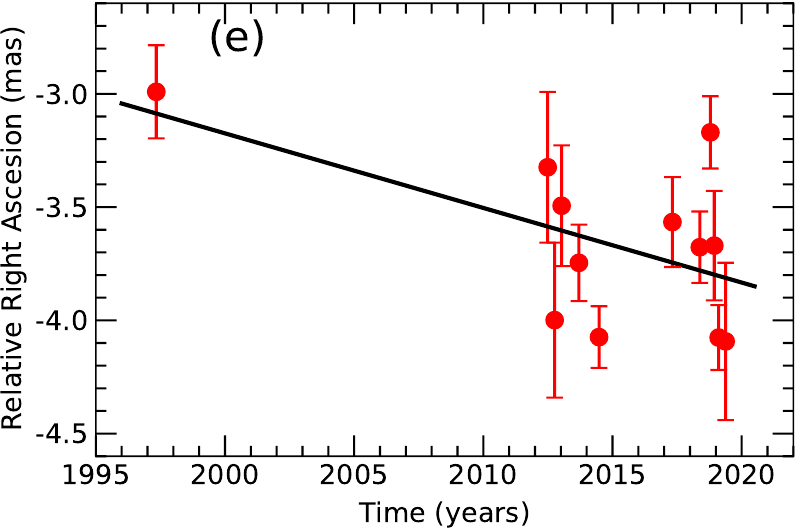}
 \includegraphics[height=4.5cm]{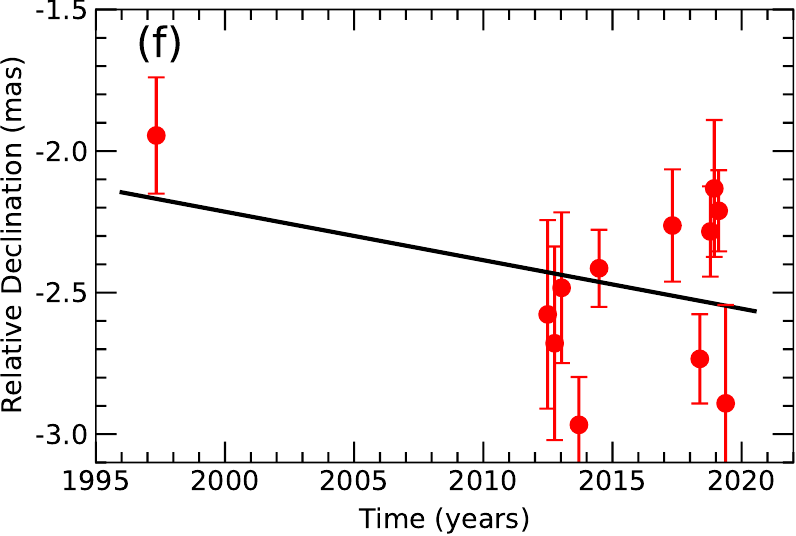}
\\
 \includegraphics[height=4.5cm]{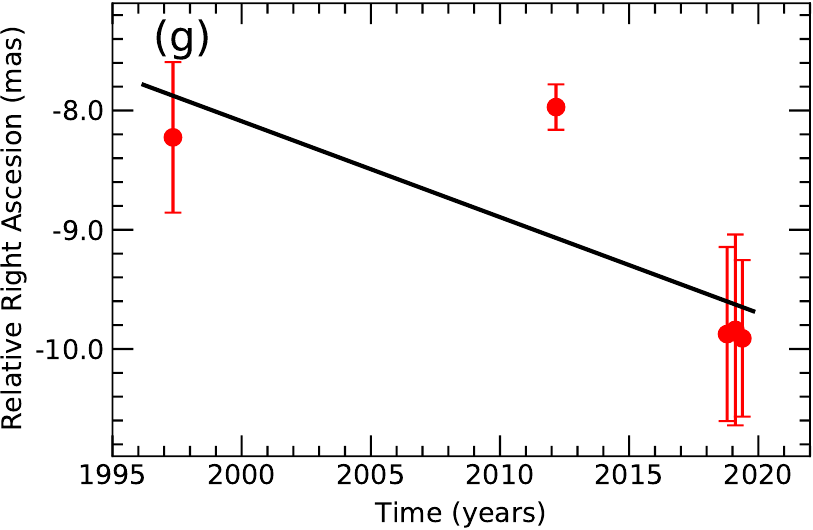}
 \includegraphics[height=4.5cm]{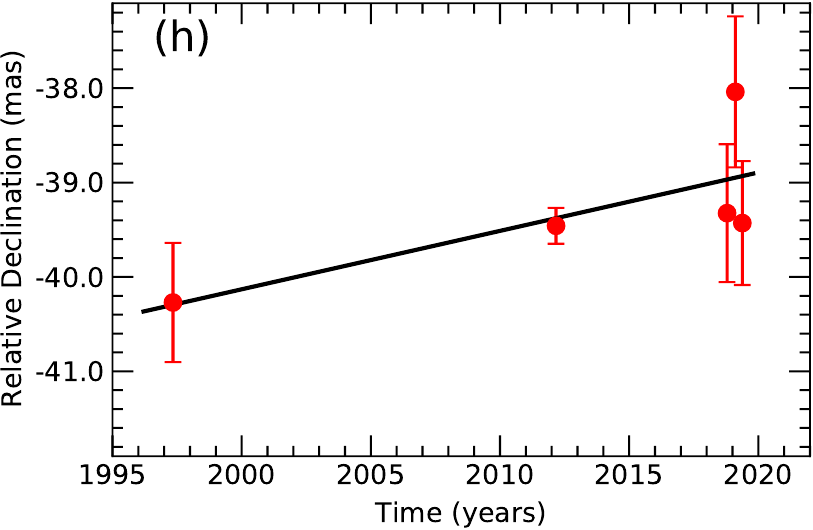}
\\
 \includegraphics[height=4.5cm]{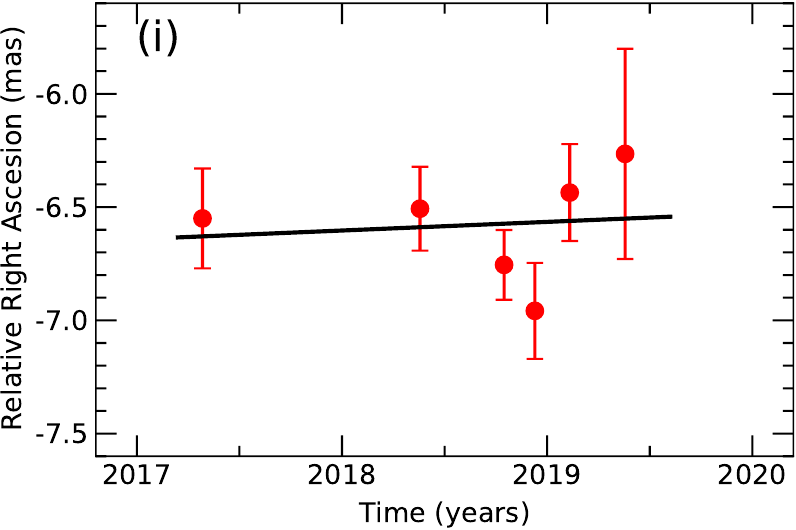}
 \includegraphics[height=4.5cm]{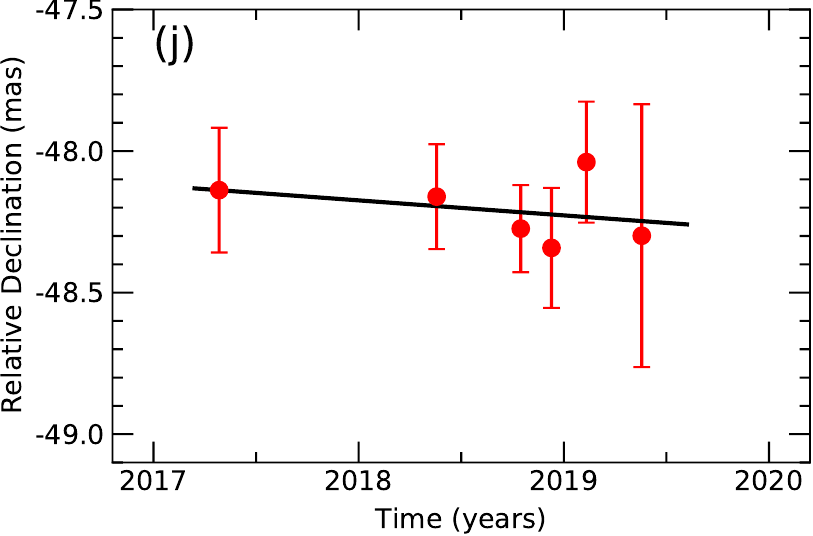}
\\
\caption{Projected RA and DEC positions of components with time for component S (a) and (b), NE (c) and (d), NW (e) and (f), SN (g) and (h), and SS (i) and (j). The solid lines represent the least-squares linear fit to the component positions as a function of time, for more details see Table \ref{tab:proper motion}.
} \label{fig:proper motion}
\end{figure*}

In order to study the velocity for the Gaussian components, we chose component N as a reference, based on its compact nature and strength.
We used a non-acceleration, two-dimensional vector fit \citep{2016AJ....152...12L} to each component's position, split into RA- and DEC-components, as a function of time.
The component SN are mostly resolved at 8 GHz because of the steep spectrum; thus, we used the data at 5 and 8 GHz together.
The data of other components (S, NE, NW, and SS) are used from the 8 GHz observations.
The fitting results and corresponding values of the angular speed, apparent speed, and angle of the motion are listed in Figure \ref{fig:proper motion} and Table \ref{tab:proper motion}.

For the northern components NE and NW, a significant motion was found, $0.017\pm0.004$ mas yr$^{-1}$ ($0.161\pm0.038\, c$) and $0.037\pm0.007$ mas yr$^{-1}$ ($0.349\pm0.066\, c$). The component NW moves in a constant direction towards the southwest direction, but the northern component NE appears to have a slight position angle change; since it is very close to the core, the model fitting of NE is much affected by the mixture with the core emission.
We note that these values rest heavily on a single observation in 1997, even though the signal-to-noise (S/N) ratio of both the NE and NW components are $>$ 50. Without it, NE and NW do not show significant positional changes with respect to N between the epochs 2012 and 2020. VLBI monitoring over an even longer time span is necessary to confirm these small changes.
For the southern components SN and SS, the jet speeds of SN and SS are $0.101\pm0.029$ mas yr$^{-1}$ ($0.954\pm0.274\, c$) and $0.064\pm0.027$ mas yr$^{-1}$ ($0.604\pm0.257\, c$). 
The jet speeds range from $0.161\, c$ to $0.954\, c$, suggesting moderate relativistic speeds.

The fitting result of component S shows no significant motion, $0.010\pm0.006$ mas yr$^{-1}$, or $0.094\pm0.057\, c$.

\subsection{Spectral Index}
\label{3.4}

The source was observed with the VLBA simultaneously at 2 and 8 GHz or 5 and 8 GHz during ten epochs spanning between 1997 and 2018. Our two quasi-simultaneous VLBA observations in 2019 were taken within 24 hours at 5 and 8 GHz.
According to the results of model fitting for these 12 epochs, listed in Table \ref{tab:model fitting}, the spectral indices are estimated and listed in column 8 of the row of the low-frequency data (2 GHz or 5 GHz) of Table \ref{tab:model fitting}.

For component N, the integrated spectral index between 2 and 8 GHz varies from $-$0.1 to $-$0.99 with a mean value of $-$0.61$\pm$0.28, and the integrated spectral index between 5 and 8 GHz varies from $-$0.07 to $-$0.48 with a mean value of $-$0.24$\pm$0.15. For component S, the integrated spectral index between 2 and 8 GHz varies from $-$0.53 to 0.20 with a mean value of $-$0.35$\pm$0.22, and the integrated spectral index between 5 and 8 GHz varies from $-$0.61 to 0.23 with a mean value of $-$0.16$\pm$0.30. 
Both components N and S, unresolved in 2 GHz images, exhibit resolved compact core and two-sided jet structures at 5 and 8 GHz.
For this reason, the spectral indices for components N and S between 2 and 8 GHz are much steeper than those between 5 and 8 GHz. We therefore believe that the spectral index between 5 and 8 GHz provides a good estimate. The mean values of N and S are $-$0.24$\pm$0.15 and $-$0.16$\pm$0.30, indicating two flat-spectrum core candidates.

For the spectral index of NE, we determined based on the quasi-simultaneous observations at 5 and 8 GHz on February 10 and 11, 2019. For the spectral index of SS, we determined based on the quasi-simultaneous observations at 5 and 8 GHz on May 20 and 21, 2019. For the spectral index of SN, we determined based on the simultaneous observations at 2 and 8 GHz on May 7, 1997.
 We find that the spectral indices of the SN, SS, and NE are $-$0.91, 1.24, and 0.65, respectively. The component NW was detected at all the 12 simultaneous or quasi-simultaneous observations. The integrated spectral index varies from $-$0.37 to $-$2.21 with a mean value of $-$0.92$\pm$0.37.
Components NW and SN show the spectral index of $-$0.92 and $-$0.91, respectively, suggesting that SN and NW are optically thin components. In contrast, an inverted spectrum was found in components SS and NE.

\begin{figure}
    \centering
    \includegraphics[width=0.45\textwidth]{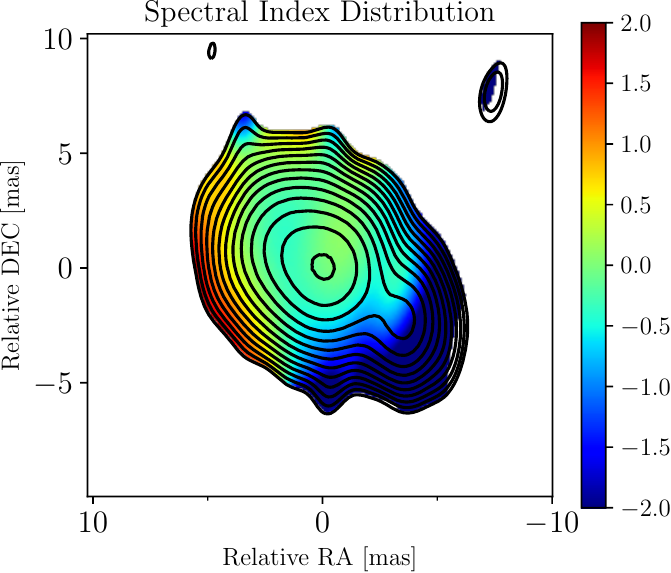}
    \includegraphics[width=0.45\textwidth]{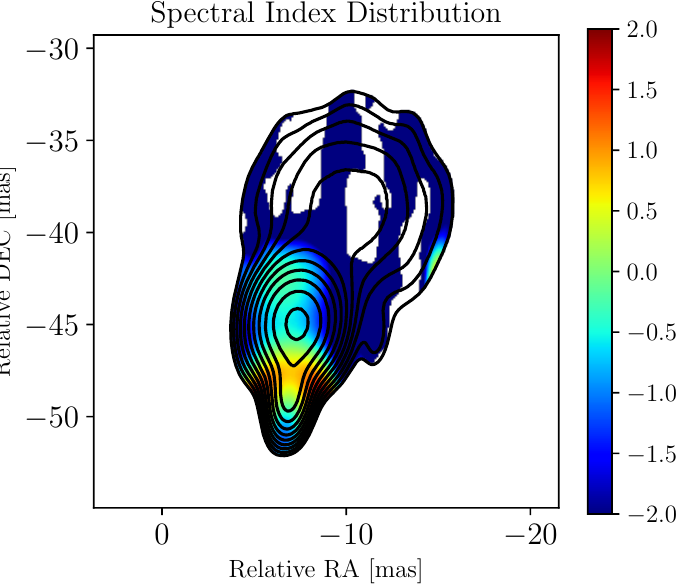}
    \\
    \caption{The spectral maps obtained by using the VLBA 4.3- and 7.6-GHz. The upper panel is for the north structure and the lower panel is for the south structure.}
    \label{fig:spix}
\end{figure}

The spectral index maps between 5 and 8 GHz are also presented in Fig.~\ref{fig:spix}.
We used the VIMAP program \citep{2014JKAS...47..195K} to align the maps on the core positions. 
The 8-GHz image was created with the same pixel size (0.1 mas) and restoring beam size as the 5-GHz image. 
Firstly, we excluded the core region with an elliptical mask about two times the size of the restoring beam. 
Then, VIMAP calculated the two-dimensional cross-correlation product to determine the shift between the two images. 
Due to the relatively close frequencies, the typical shift is about 0.05 mas due to the frequency-dependent opacity effect. 
After correcting this relative offset, we computed the spectral index map and superimposed it (in colored scale) on the 5 GHz total intensity contours.

For the northern complex structure (upper panel of Fig. \ref{fig:spix}), the spectral-index distribution shows that the N component has a flat spectrum ($\alpha$ $\sim$ 0), characteristic of a synchrotron self-absorbed core, while the NE and NW components have an inverted and steep spectrum expected for an optically thin jet feature.
This is evidence of a source with a core and two-sided jet structure \citep{2021MNRAS.506.1609C,2023Galax..11...42C}.
For the southern complex structure (lower panel of Fig. \ref{fig:spix}), the spectral-index distribution shows a similar result that the S component has a flat spectrum ($\alpha$ $\sim$ 0), characteristic of a synchrotron self-absorbed core, while the SS and SN components have an inverted and steep spectrum expected for an optically thin jet feature.

\section{Discussion}
\label{4}

Based on the radio morphology, variability, proper motions, and spectral index of J1543$-$0757 revealed by multifrequency VLBI observations, there are three scenarios to explain the source's physical nature: gravitational lensing, a compact symmetric object (CSO), or a dual SMBH system.

The first possibility is gravitational lensing. The two components N and S we observed in J1543$-$0757 are gravitationally lensed by a foreground compact object. Since the first gravitational lens was discovered by \citet {1979Natur.279..381W}, there are under 100 lensed sources known at radio frequency with a maximum image separation between 0.3 and 6 arcsec \citep[e.g.,][]{1999MNRAS.307..225K,2003MNRAS.341...13B,2003MNRAS.341....1M,2024MNRAS.530..221J}. Recently, several VLBI observations have been used in search for milliarcsecond-lenses with image separations $>$ 1.5 mas \citep[e.g.,][]{2001PhRvL..86..584W,2003MNRAS.341...13B,2019MNRAS.483.2125S,2021MNRAS.507L...6C},  but found none were confirmed as milliarcsecond-lenses. To date, the smallest known gravitational lens is B0218+357 with a separation of 334 mas \citep{1992AJ....104.1320O,1993MNRAS.261..435P}. If J1543$-$0757 is gravitational lensed, this will be the first confirmed milliarcsecond-lens within 50 mas. However, we note that the light curves of components N and S are very different (see Fig. \ref{fig:lightcurve}). Although different light ray paths and gravitational delays cause the emission to arrive at different times in various images, we would expect them to be proportional, with a short time delay. Given the short time delays between images in milliarcsecond-lenses \citep[e.g. around 11 days for B0218+357,][]{2018MNRAS.476.5393B}, we attempted to apply a short time delay (in step of 1 day) ranging from 1 to 30 days, comparing each point with its corresponding counterpart in the other image using interpolation. However, we did not find any cross-correlation in the light-curve data between components N and S. Based on the known rarity of compact, lensed sources and significant difference in the light curves of components N and S of J1543$-$0757, we eliminate gravitational lensing as a potential mechanism to describe the observed parsec-scale morphology.

The second scenario is a CSO. The deficit of extended radio emission \citep{1998AJ....115.1693C}, slow-moving emission, and steep spectrum \citep{2004A&A...424...91E} are generally consistent with the CSO classification \citep{1994ApJ...425..568C,1996ApJ...460..612R,2024ApJ...961..240K}. When observed with sufficient sensitivity and resolution, the overall structure of a CSO can appear edge-brightened, that is, the emission is brightest at the CSO’s extremities, called hotspots, and becomes fainter closer to its central SMBH \citep{2012ApJS..198....5A,2014ApJ...792L...8W,2024ApJ...961..241K,2024ApJ...961..242R}. The large separation of the N and S components suggests a CSO-like type morphology at low frequencies. However, compared to the typical CSOs \citep[e.g.,][]{2011A&A...535A..24S,2012ApJ...760...77A}, J1543$-$0757 is much different, having two flat spectrum components and two-sided jet structures in 5 and 8 GHz VLBI images. 

To further confirm this, we compared the Gaia optical coordinates with radio positions obtained from our VLBI observations.  Since a double radio source without a firmly detected VLBI core, but with the Gaia optical position and falling in between the two opposite knots/hotspots, is a strong indication for a CSO \citep{2020MNRAS.496.1811K}. From Fig. \ref{fig:image}, we found the radio position of component N is much more consistent with the Gaia position with only 4.3 mas offset. This provides further evidence for the radio core of N which is more related to a SMBH. We also found that the Gaia has a very large astrometric excess noise with about 60 mas \citep{2020yCat.1350....0G}, suggesting a candidate dual/off nucleus quasars at parsec scales \citep{2023ApJ...958...29C}.

\citet{1997ApJ...485..112B} predicted that the host galaxies of compact and young AGNs might harbor a gas-rich medium. J1543$-$0757 is classified as a GPS source, but the non-detection of HI absorption line in the Giant Metrewave Radio Telescope observations \citep{2018MNRAS.473...59A} implies low column density of gas and dust. We further exclude the scenario that N and S are the jet components lit up in a collision with a dense interstellar medium.

The last possibility is that the two observed compact, flat-spectrum components are part of a dual SMBHs system formed from a merger. In this case, components N and S (radio core candidates) would be interpreted as active nuclei. Several of the observed features, such as two-sided jet morphology both in the north and south features and the spectrum, are highly supportive of this scenario. 

We assume that the current dual SMBHs hypothesis is correct. The assumption that the two SMBHs are in N and S and that the expected distance is $\sim$46 mas or 130 pc (between N and S) should be validated at higher simultaneous dual-frequency frequencies. VLBI monitoring observations at mm wavelengths to search for new components from the two candidate cores N and S to see if a trend existed in which components appeared to emerge from the candidate BH and move away. This would go a long way towards proving this scenario. Simultaneous observations with the VLBI are constructive to confirm this. The VLBI 15 and 22 GHz observations are important to detect the compact components and resolve the possible jet structure. We can test the compactness of the two radio cores, if one or two are the jet hotspots, this should be a core-jet or an edge-brightened double-jet structure \citep{2014ApJ...792L...8W}. The spectra of the two components at high frequencies can also constrain testing CSO or pc-scale dual SMBHs.

\section{Summary}
\label{5}

We have reported on the high-resolution VLBA and EVN observations of GPS source J1543$-$0757 at 1.6, 2, 5, and 8 GHz. The results reveal two flat spectrum radio cores with a separation of $\sim$46 mas ($\sim$130 pc). The significant difference in the light curves of components N and S eliminates lensing as a mechanism to describe the observed parsec-scale morphology. The CSO scenario was also ruled out because of the flat spectrum component and two-sided jet structures both in N and S at 5 and 8 GHz. The optical Gaia position is consistent with component N and does not fall in between the two opposite components N and S, further excluding the hypothesis. The most possible scenario is a dual SMBHs system. Higher resolution and frequency VLBI imaging could in principle prove that the sources are more compact, ruling out the explanation involving hotspots in CSO in the future.

In summary, J1543 $-$ 0757 will provide an important test of models for galaxy merging and dual SMBHs formation. Further high-angular resolution and frequency observations will be needed to understand the nature of these two components and the origin of their radio emission.

\section*{Acknowledgements}

We thank the referee for the helpful comments which improve the manuscript.This work was supported by Brain Pool Program through the National Research Foundation of Korea (NRF) funded by the Ministry of Science and ICT (2019H1D3A1A01102564, RS-2024-00407499). 
The European VLBI Network (EVN) is a joint facility of independent European, African, Asian, and North American radio astronomy institutes. Scientific results from data presented in this publication. 
The VLBA observations were sponsored by Shanghai Astronomical Observatory through an MoU with the NRAO (Project code: BT146). 
The Very Long Baseline Array is a facility of the National Science Foundation operated under cooperative agreement by Associated Universities, Inc. This work made use of the DiFX software correlator developed at Swinburne University of Technology as part of the Australian Major National Research Facilities program.

\section*{Data Availability}
The correlation data used in this article are available in the EVN data archive (\url{http://www.jive.nl/select-experiment}) and VLBA data archive (\url{https://archive.nrao.edu/archive/archiveproject.jsp}). The calibrated visibility data underlying this article can be requested from the corresponding authors.




\bibliography{sample631}{}
\bibliographystyle{aasjournal}




\begin{table*}
\centering
    \caption{Description of the archival data used in this paper. Columns give (1) observing frequency, (2) total flux density, (3) uncertainty on the flux density, (4) references for data points: (1) GaLactic and Extragalactic All-sky MWA (GLEAM) at 72 - 231 MHz \citep{2017MNRAS.464.1146H}, (2) Texas survey at 365 MHz \citep{1996AJ....111.1945D}, (3) Molonglo Reference Catalogue \citep[MRC;][]{1981MNRAS.194..693L} at 408 MHz, (4) NVSS at 1.40 GHz \citep{1998AJ....115.1693C}, (5) ATCA at 1.4, 2.5, 4.8 and 8.6 GHz \citep{2003PASJ...55..351T}, (6) Parkes-MIT-NRAO (PMN) surveys at 4.85 GHz \citep{1995ApJS...97..347G}, (7) VSOP 5 GHz Active Galactic Nucleus Survey at 5 GHz \citep{2004ApJ...616..110H}, (8) Combined Radio All-Sky Targeted Eight GHz Survey (CRATES) at 8.4 GHz \citep{2007ApJS..171...61H}, (9) Australia Telescope 20 GHz Survey (AT20G) at 20 GHz \citep{2010MNRAS.402.2403M}, and (10) Korean VLBI Network Calibrator Survey (KVNCS) at 22 GHz \citep{2017ApJS..228...22L}. }
	\begin{tabular}{cccc}
	\hline\hline
Frequency   & S    & $\rm \sigma$     & reference \\
 (GHz)      & (Jy) & (Jy beam$^{-1}$) &           \\
\hline
0.076 &	0.055 &	0.004 & 1 \\
0.084 &	0.082 &	0.007 & 1 \\
0.092 &	0.048 &	0.004 & 1 \\
0.099 &	0.129 &	0.010 & 1 \\
0.107 &	0.237 &	0.019 & 1 \\
0.115 &	0.069 &	0.006 & 1 \\
0.122 &	0.176 &	0.014 & 1 \\
0.130 &	0.161 &	0.013 & 1 \\
0.143 &	0.244 &	0.020 & 1 \\
0.151 &	0.323 &	0.026 & 1 \\
0.158 &	0.335 &	0.027 & 1 \\
0.166 &	0.366 &	0.029 & 1 \\
0.174 &	0.426 &	0.034 & 1 \\
0.181 &	0.441 &	0.035 & 1 \\
0.189 &	0.397 &	0.032 & 1 \\
0.197 &	0.394 &	0.032 & 1 \\
0.204 &	0.489 &	0.039 & 1 \\
0.212 &	0.493 &	0.039 & 1 \\
0.220 &	0.528 &	0.042 & 1 \\
0.227 &	0.539 &	0.043 & 1 \\
0.365 &	1.346 &	0.057 & 2 \\
0.408 &	1.650 &	0.050 & 3 \\
1.400 &	1.509 &	0.053 & 4 \\
1.400 &	1.540 &	0.077 & 5 \\
2.500 &	1.330 &	0.067 & 5 \\
4.800 &	1.000 &	0.050 & 5 \\
4.850 &	1.013 &	0.054 & 6 \\
5.000 &	1.000 &	0.100 & 7 \\
8.440 &	0.717 &	0.036 & 8 \\
8.600 &	0.690 &	0.035 & 5 \\
19.900& 0.385 &	0.019 & 9 \\
21.700& 0.310 &	0.030 & 10\\	
\hline
	\end{tabular}
\label{tab:SED}
\begin{flushleft}
\end{flushleft}
\end{table*}

\begin{table*}
\centering
    \caption{Log of VLBI Observations. Columns give (1) date, (2) observing frequency, (3) project code, (4) total bandwidth, (5) integration time, (6) polarization, (7) map specific intensity, (8) integrated flux density, (9) references: (1) $ \rm http://astrogeo.org/ $, (2) EVN observation (PI: H.E. Bignall), calibrated data downloaded from the EVN data archive (http://archive.jive.nl/scripts/portal.php), (3) \citet{2022ApJ...934...39S} }
	\begin{tabular}{ccccccccc}
	\hline\hline
Date        & Frequency   & code   & Bandwidth & Time      & Pol. & $\rm S_{Peak}$  & $\rm S_{int}$ & reference \\
            & (GHz)       &        & (MHz)     & (minutes) &      &(Jy beam$^{-1}$) &        (Jy)    &          \\
\hline
1997 May 07 &  2.27       & BB023  & 32        & 5         & 1    & 0.536           &     1.571      &    1     \\
            &  8.34       & BB023  & 32        & 5         & 1    & 0.039           &     0.813      &    1     \\
2005 Jun 03 &  1.66       & EB030  & 16        & 29        & 4    & 0.824           &     1.574      &    2     \\
2012 Feb 29 &  8.36       & BC201AL& 128       & 15        & 1    & 0.081           &     0.651      &    1     \\
2012 Jue 28 &  2.31       & RDV93  & 64        & 11        & 1    & 0.679           &     1.382      &    1     \\
            &  8.64       & RDV93  & 64        & 11        & 1    & 0.271           &     0.698      &    1     \\
2012 Oct 03 &  2.31       & RDV95  & 32        & 12        & 1    & 0.733           &     1.441      &    1     \\
            &  8.64       & RDV95  & 32        & 12        & 1    & 0.204           &     0.642      &    1     \\
2013 Jan 09 &  2.31       & RDV97  & 32        & 12        & 1    & 0.436           &     1.246      &    1     \\
            &  8.65       & RDV97  & 32        & 12        & 1    & 0.055           &     0.425      &    1     \\
2013 Sep 11 &  2.31       & RV101  & 32        & 11        & 1    & 0.491           &     1.404      &    1     \\
            &  8.64       & RV101  & 32        & 11        & 1    & 0.129           &     0.668      &    1     \\
2014 Jun 26 &  2.31       & RV105  & 32        & 11        & 1    & 0.389           &     1.149      &    1     \\
            &  8.64       & RV105  & 32        & 11        & 1    & 0.166           &     0.658      &    1     \\
2017 Apr 29 &  2.29       & UF001G & 128       & 6         & 1    & 0.476           &     1.347      &    1     \\
            &  8.67       & UF001G & 128       & 6         & 1    & 0.048           &     0.381      &    1     \\
2018 May 20 &  2.25       & UG002H & 96        & 6         & 1    & 0.469           &     1.427      &    1     \\
            &  8.65       & UG002H & 96        & 6         & 1    & 0.057           &     0.450      &    1     \\
2018 Oct 16 &  4.34       & SB072B0& 256       & 7         & 1    & 0.110           &     0.759      &    1     \\
            &  7.62       & SB072B0& 256       & 7         & 1    & 0.046           &     0.604      &    1     \\
2018 Dec 11 &  2.28       & BP222B2& 256       & 5         & 1    & 0.423           &     1.383      &    1     \\
            &  8.65       & BP222B2& 256       & 5         & 1    & 0.027           &     0.660      &    1     \\
2019 Feb 10 &  4.87       & BT146C1& 256       & 20        & 2    & 0.168           &     0.859      &    3     \\
2019 Feb 11 &  8.37       & BT146X1& 256       & 21        & 2    & 0.045           &     0.701      &    3     \\
2019 May 20 &  4.87       & BT146C2& 256       & 20        & 2    & 0.130           &     0.845      &    3     \\
2019 May 21 &  8.37       & BT146X2& 256       & 21        & 2    & 0.024           &     0.652      &    3     \\
\hline
	\end{tabular}
\label{tab:observation log}
\begin{flushleft}
\end{flushleft}
\end{table*}

\begin{table*}
\centering
\renewcommand{\tabcolsep}{3.5mm}
    \caption{Parameters of the fitted Gaussian model components. Columns give (1) Epoch, (2) component label, (3) observing frequency, (4) integrated flux density, (5) - (6) distance and position angle to the compact core component N, (7) size of the deconvolved Gaussian model, (8) spectral index}
	\begin{tabular}{cccccccc}
	\hline\hline
Date        & Component   & Frequency   & S        & R      & PA        & $\theta$  & $\alpha$ \\
            &             &     (GHz)   & (Jy)     & (mas)  &  (deg)    &(mas)      &          \\
\hline
1997 May 07 & N  & 2.27 & 0.705$\pm$0.071 & 0.0                & 0.0     & 3.778$\pm$0.378   & $-$0.69 \\
            &    & 8.34 & 0.284$\pm$0.028 & 0.0                & 0.0     & 2.640$\pm$0.264   &         \\
            & S  & 2.27 & 0.245$\pm$0.025 & 46.121$\pm$0.431   & $-$169.84 & 2.875$\pm$0.288 & $-$0.37 \\
            &    & 8.34 & 0.151$\pm$0.015 & 45.743$\pm$0.469   & $-$171.57 & 3.126$\pm$0.313 &         \\
            & NE & 8.34 & 0.191$\pm$0.019 & 2.047$\pm$0.359    & 70.21     & 2.396$\pm$0.240 &         \\
            & NW & 2.27 & 0.344$\pm$0.034 & 3.568$\pm$0.613    & $-$123.93 & 4.085$\pm$0.409 & $-$0.90 \\
            &    & 8.34 & 0.107$\pm$0.011 & 3.568$\pm$0.308    & $-$123.04 & 2.055$\pm$0.206 &         \\
            & SN & 2.27 & 0.253$\pm$0.025 & 40.530$\pm$1.146   & $-$163.74 & 7.640$\pm$0.764 & $-$0.91 \\
            &    & 8.34 & 0.078$\pm$0.008 & 41.102$\pm$0.948   & $-$168.46 & 6.318$\pm$0.632 &         \\
2005 Jun 03 & N  & 1.66 & 1.090$\pm$0.109 & 0.0                & 0.0       & 3.850$\pm$0.385 &         \\
            & S  & 1.66 & 0.446$\pm$0.045 & 42.568$\pm$0.605   & -169.41   & 4.034$\pm$0.403 &         \\
2012 Feb 29 & N  & 8.36 & 0.280$\pm$0.028 & 0.0                & 0.0       & 3.118$\pm$0.312 &         \\
            & S  & 8.36 & 0.155$\pm$0.016 & 46.188$\pm$0.514   & $-$171.89 & 3.429$\pm$0.343 &         \\
            & NE & 8.36 & 0.053$\pm$0.005 & 2.581$\pm$0.137    & 51.638    & 0.913$\pm$0.091 &         \\
            & SN & 8.36 & 0.011$\pm$0.001 & 40.255$\pm$0.285   & $-$168.58 & 1.901$\pm$0.190 &         \\
2012 Jue 28 & N  & 2.31 & 0.749$\pm$0.075 & 0.0                & 0.0       & 4.195$\pm$0.420 & $-$0.45 \\
            &    & 8.64 & 0.411$\pm$0.041 & 0.0                & 0.0       & 3.130$\pm$0.313 &         \\
            & S  & 2.31 & 0.113$\pm$0.011 & 47.076$\pm$0.335   & $-$171.11 & 2.231$\pm$0.223 & 0.20    \\
            &    & 8.64 & 0.148$\pm$0.015 & 45.743$\pm$0.469   & $-$171.57 & 3.126$\pm$0.313 &         \\
            & NW & 2.31 & 0.183$\pm$0.018 & 4.209$\pm$0.480    & $-$125.23 & 3.203$\pm$0.320 & $-$0.37 \\
            &    & 8.64 & 0.113$\pm$0.011 & 3.568$\pm$0.308    & $-$123.04 & 2.055$\pm$0.206 &         \\
2012 Oct 03 & N  & 2.31 & 0.759$\pm$0.076 & 0.0                & 0.0       & 4.121$\pm$0.412 & $-$0.53 \\
            &    & 8.64 & 0.373$\pm$0.037 & 0.0                & 0.0       & 3.227$\pm$0.323 &         \\
            & S  & 2.31 & 0.266$\pm$0.027 & 46.121$\pm$0.431   & $-$169.84 & 2.875$\pm$0.288 & $-$0.36 \\
            &    & 8.64 & 0.166$\pm$0.017 & 45.442$\pm$0.692   & $-$171.46 & 4.611$\pm$0.461 &         \\
            & NW & 2.31 & 0.192$\pm$0.019 & 4.139$\pm$0.445    & $-$122.93 & 2.964$\pm$0.296 & $-$0.64 \\
            &    & 8.64 & 0.082$\pm$0.008 & 4.813$\pm$0.309    & $-$123.82 & 2.058$\pm$0.206 &         \\
2013 Jan 09 & N  & 2.31 & 0.690$\pm$0.069 & 0.0                & 0.0       & 4.082$\pm$0.408 & $-$0.99 \\
            &    & 8.65 & 0.187$\pm$0.019 & 0.0                & 0.0       & 1.714$\pm$0.171 &         \\
            & S  & 2.31 & 0.179$\pm$0.018 & 46.174$\pm$0.612   & $-$170.86 & 4.082$\pm$0.408 & $-$0.42 \\
            &    & 8.65 & 0.102$\pm$0.010 & 45.668$\pm$0.468   & $-$171.32 & 3.119$\pm$0.312 &         \\
            & NE & 8.65 & 0.063$\pm$0.006 & 2.394$\pm$0.216    & 88.06     & 1.442$\pm$0.144 &         \\
            & NW & 2.31 & 0.132$\pm$0.013 & 4.413$\pm$0.408    & $-$122.97 & 2.721$\pm$0.272 & $-$0.44 \\
            &    & 8.65 & 0.073$\pm$0.007 & 4.286$\pm$0.399    & $-$125.39 & 2.663$\pm$0.266 &         \\
2013 Sep 11 & N  & 2.31 & 0.746$\pm$0.075 & 0.0                & 0.0       & 4.180$\pm$0.418 & $-$0.38 \\
            &    & 8.64 & 0.448$\pm$0.045 & 0.0                & 0.0       & 4.019$\pm$0.402 &         \\
            & S  & 2.31 & 0.162$\pm$0.016 & 46.633$\pm$0.369  & $-$171.18  & 2.457$\pm$0.246 & $-$0.14 \\
            &    & 8.64 & 0.134$\pm$0.013 & 45.816$\pm$0.486  & $-$171.58  & 3.242$\pm$0.324 &         \\
            & NW & 2.31 & 0.208$\pm$0.021 & 4.103$\pm$0.539   & $-$125.31  & 3.594$\pm$0.359 & $-$1.07 \\
            &    & 8.64 & 0.051$\pm$0.005 & 4.779$\pm$0.254   & $-$128.38  & 1.689$\pm$0.169 &         \\
2014 Jun 26 & N  & 2.31 & 0.615$\pm$0.062 & 0.0               & 0.0        & 4.145$\pm$0.415 & $-$0.10 \\
            &    & 8.64 & 0.536$\pm$0.054 & 0.0               & 0.0        & 3.561$\pm$0.356 &         \\
            & S  & 2.31 & 0.171$\pm$0.017 & 46.354$\pm$0.887  & $-$171.18  & 5.916$\pm$0.592 & $-$0.53 \\
            &    & 8.64 & 0.085$\pm$0.009 & 46.122$\pm$0.479  & $-$171.27  & 3.191$\pm$0.319 &         \\
            & NW & 2.31 & 0.168$\pm$0.017 & 4.318$\pm$0.559   & $-$125.03  & 3.726$\pm$.0373 & $-$0.73 \\
            &    & 8.64 & 0.064$\pm$0.006 & 4.736$\pm$0.204   & $-$120.65  & 1.361$\pm$0.136 &         \\
2017 Apr 29 & N  & 2.31 & 0.382$\pm$0.038 & 0.0               & 0.0        & 3.545$\pm$0.355 & $-$0.72 \\
            &    & 8.67 & 0.146$\pm$0.015 & 0.0               & 0.0        & 3.561$\pm$0.356 &         \\
            & S  & 2.31 & 0.112$\pm$0.011 & 46.727$\pm$0.556  & $-$171.67  & 3.709$\pm$0.371 & $-$0.50 \\
            &    & 8.67 & 0.058$\pm$0.006 & 45.220$\pm$0.479  & $-$171.31  & 3.191$\pm$0.319 &         \\

\end{tabular}
\label{tab:model fitting}
\end{table*}

\begin{table*}
%
\medskip
\centering
\renewcommand{\tabcolsep}{3.5mm}
\begin{tabular}{cccccccc}
\hline\hline
Date        & Component   & Frequency   & S        & R      & PA        & $\theta$  & $\alpha$ \\
            &             &     (GHz)   & (Jy)     & (mas)  &  (deg)    &(mas)      &          \\
\hline
            & NE & 8.67 & 0.078$\pm$0.008 & 2.314$\pm$0.262   & 91.08      & 1.741$\pm$0.174 &         \\
            & NW & 2.31 & 0.135$\pm$0.014 & 4.576$\pm$0.437   & $-$121.53  & 2.914$\pm$0.291 & $-$0.60 \\
            &    & 8.67 & 0.061$\pm$0.006 & 4.223$\pm$0.298   & $-$122.40  & 1.984$\pm$0.198 &         \\
            & SS & 8.67 & 0.040$\pm$0.004 & 48.582$\pm$0.331  & $-$172.25  & 2.204$\pm$0.220 &         \\
2018 May 20 & N  & 2.25 & 0.800$\pm$0.080 & 0.0               & 0.0        & 4.456$\pm$0.446 & $-$0.98 \\
            &    & 8.65 & 0.213$\pm$0.021 & 0.0               & 0.0        & 2.412$\pm$0.241 &         \\
            & S  & 2.25 & 0.122$\pm$0.012 & 44.928$\pm$0.335  & $-$170.02  & 2.235$\pm$0.224 & $-$0.43 \\
            &    & 8.65 & 0.069$\pm$0.007 & 45.506$\pm$0.444  & $-$171.24  & 2.959$\pm$0.296 &         \\
            & NE & 8.65 & 0.082$\pm$0.008 & 2.297$\pm$0.441   & 86.56      & 2.943$\pm$0.294 &         \\
            & NW & 2.25 & 0.128$\pm$0.013 & 4.714$\pm$0.453   & $-$120.68  & 3.023$\pm$0.302 & $-$0.97 \\
            &    & 8.65 & 0.044$\pm$0.004 & 4.582$\pm$0.234   & $-$126.63  & 1.577$\pm$0.158 &         \\
            & SS & 8.65 & 0.041$\pm$0.004 & 48.599$\pm$0.278 & $-$172.31 & 1.851$\pm$0.185 &         \\
2018 Oct 16 & N  & 4.34 & 0.417$\pm$0.042 & 0.0              & 0.0       & 3.837$\pm$0.384 & $-$0.15 \\
            &    & 7.62 & 0.382$\pm$0.038 & 0.0              & 0.0       & 3.469$\pm$0.347 &         \\
            & S  & 4.34 & 0.113$\pm$0.011 & 45.758$\pm$0.419 & $-$170.56 & 2.790$\pm$0.279 & $-$0.61 \\
            &    & 7.62 & 0.165$\pm$0.017 & 46.200$\pm$0.577 & $-$171.26 & 3.845$\pm$0.385 &         \\
            & NE & 4.34 & 0.016$\pm$0.002 & 3.850$\pm$0.195  & 33.25     & 1.298$\pm$0.130 &         \\
            & NW & 4.34 & 0.098$\pm$0.010 & 4.873$\pm$0.473  & $-$123.82 & 3.154$\pm$0.315 & $-$0.97 \\
            &    & 7.62 & 0.056$\pm$0.006 & 3.907$\pm$0.239  & $-$125.77 & 1.595$\pm$0.160 &         \\
            & SN & 4.34 & 0.011$\pm$0.001 & 40.255$\pm$0.285 & $-$168.58 & 1.901$\pm$0.190 &         \\
            & SS & 4.34 & 0.026$\pm$0.003 & 48.744$\pm$0.231 & $-$172.03 & 1.539$\pm$0.154 &         \\
2018 Dec 11 & N  & 2.28 & 0.770$\pm$0.077 & 0.0              & 0.0       & 4.320$\pm$0.432 & $-$0.48 \\
            &    & 8.65 & 0.400$\pm$0.020 & 0.0              & 0.0       & 3.613$\pm$0.361 &         \\
            & S  & 2.28 & 0.137$\pm$0.014 & 46.419$\pm$0.368 & $-$170.81 & 2.450$\pm$0.245 & $-$0.12 \\
            &    & 8.65 & 0.116$\pm$0.012 & 45.186$\pm$0.437 & $-$171.41 & 2.912$\pm$0.291 &         \\
            & NW & 2.28 & 0.217$\pm$0.022 & 4.460$\pm$0.633  & $-$123.82 & 4.218$\pm$0.413 & $-$0.79 \\
            &    & 8.65 & 0.076$\pm$0.008 & 4.244$\pm$0.363  & $-$120.15 & 2.418$\pm$0.242 &         \\
            & SS & 8.65 & 0.066$\pm$0.007 & 48.840$\pm$0.318 & $-$171.81 & 2.118$\pm$0.212 &         \\
2019 Feb 10 & N  & 4.87 & 0.402$\pm$0.040 & 0.0              & 0.0       & 3.685$\pm$0.369 & $-$0.07 \\
            & S  & 4.87 & 0.122$\pm$0.012 & 45.116$\pm$0.404 & $-$170.71 & 2.692$\pm$0.269 &    0.23 \\
            & NE & 4.87 & 0.095$\pm$0.010 & 2.610$\pm$0.510  & 57.78     & 3.402$\pm$0.340 &    0.65 \\
            & NW & 4.87 & 0.097$\pm$0.010 & 4.612$\pm$0.419  & $-$125.86 & 2.795$\pm$0.280 & $-$2.21 \\
            & SN & 4.87 & 0.104$\pm$0.010 & 39.291$\pm$1.201 & $-$165.50 & 8.006$\pm$0.800 &         \\
            & SS & 4.87 & 0.037$\pm$0.004 & 48.468$\pm$0.321 & $-$172.48 & 2.138$\pm$0.214 &         \\
2019 Feb 11 & N  & 8.37 & 0.389$\pm$0.039 & 0.0              & 0.0       & 3.505$\pm$0.351 &         \\
            & NE & 8.37 & 0.138$\pm$0.014 & 2.615$\pm$0.387  & 62.14     & 2.579$\pm$0.258 &         \\
            & S  & 8.37 & 0.145$\pm$0.015 & 45.494$\pm$0.308 & $-$171.52 & 2.054$\pm$0.205 &         \\
            & NW & 8.37 & 0.029$\pm$0.003 & 4.637$\pm$0.215  & $-$138.47 & 1.431$\pm$0.143 &         \\
2019 May 20 & N  & 4.87 & 0.484$\pm$0.048 & 0.0              & 0.0       & 4.073$\pm$0.407 & $-$0.27 \\
            & S  & 4.87 & 0.135$\pm$0.014 & 45.879$\pm$0.381 & $-$170.57 & 2.542$\pm$0.254 & $-$0.13 \\
            & NW & 4.87 & 0.104$\pm$0.010 & 5.010$\pm$0.533  & $-$125.24 & 3.551$\pm$0.355 & $-$1.35 \\
            & SN & 4.87 & 0.101$\pm$0.010 & 40.656$\pm$0.985 & $-$165.89 & 6.566$\pm$0.656 &         \\
            & SS & 4.87 & 0.021$\pm$0.002 & 49.695$\pm$0.222 & $-$172.76 & 1.481$\pm$0.148 & 1.24    \\
2019 May 21 & N  & 8.37 & 0.438$\pm$0.044 & 0.0              & 0.0       & 4.457$\pm$0.446 &         \\
            & S  & 8.37 & 0.123$\pm$0.012 & 46.269$\pm$0.475 & $-$171.57 & 3.166$\pm$0.317 &         \\
            & NW & 8.37 & 0.050$\pm$0.005 & 4.633$\pm$0.521  & $-$118.86 & 3.472$\pm$0.347 &         \\
            & SS & 8.37 & 0.041$\pm$0.004 & 48.704$\pm$0.696 & $-$172.61 & 4.643$\pm$0.464 &         \\
\hline
	\end{tabular}
\begin{flushleft}
\end{flushleft}
\end{table*}

\begin{table*}
\centering
    \caption{Proper Motions of the Components. Columns give: (1) component; (2)-(3) velocity in RA and DEC direction; (4)-(5) apparent speed of the component; (6) angles measured from north through east.}
	\begin{tabular}{cccccc}
	\hline\hline
Component & Velocity in RA   & Velocity in DEC & Velocity        & Velocity        & Angle of Motion \\
          & (mas yr$^{-1}$)  & (mas yr$^{-1}$) & (mas yr$^{-1}$) &    (c)          &    (deg)        \\
\hline
S         & $-$0.009$\pm$0.002 & 0.006$\pm$0.005    & 0.010$\pm$0.006  & 0.094$\pm$0.057 & $-$56.3  \\
NE        & 0.016$\pm$0.004    & 0.005$\pm$0.001    & 0.017$\pm$0.004  & 0.161$\pm$0.038 & 72.6     \\
NW        & $-$0.033$\pm$0.006 & $-$0.017$\pm$0.003 & 0.037$\pm$0.007  & 0.349$\pm$0.066 & $-$117.3 \\
SN        & $-$0.081$\pm$0.015 & 0.061$\pm$0.025    & 0.101$\pm$0.029  & 0.954$\pm$0.274 & $-$53.1  \\
SS        &    0.038$\pm$0.008 & $-$0.052$\pm$0.026 & 0.064$\pm$0.027  & 0.604$\pm$0.257 & 143.8    \\
\hline
	\end{tabular}
\label{tab:proper motion}
\begin{flushleft}
\end{flushleft}
\end{table*}

\vspace{5mm}
\facilities{HST(STIS), Swift(XRT and UVOT), AAVSO, CTIO:1.3m,
CTIO:1.5m,CXO}


\software{astropy \citep{2013A&A...558A..33A,2018AJ....156..123A},  
          Cloudy \citep{2013RMxAA..49..137F}, 
          Source Extractor \citep{1996A&AS..117..393B}
          }

\end{document}